\documentclass[twocolumn,showpacs]{revtex4}

\usepackage{epsf}

\begin{document}
%\draft
\preprint{}
%%%%%%%%%%%%%%%%%%%%%%%%%%%%%%%%%%%%%%%%%%%%%%%%%%%%%%%%%
\title{Magnetization process for a quasi-one-dimensional 
$S=1$ antiferromagnet
}
%%%%%%%%%%%%%%%%%%%%%%%%%%%%%%%%%%%%%%%%%%%%%%%%%%%%%%
\author{Akira Kawaguchi, Akihisa Koga,
Kouichi Okunishi$^1$ and Norio Kawakami}
\address{Department of Applied Physics,
Osaka University, Suita, Osaka 565-0871, Japan \\ 
$^1$Department of Physics, Niigata University, 
Igarashi 2, Niigata 950-2181 Japan}
\date{\today}
%%\maketitle
%----------------------------------------------------------------------
%                              Abstract
%----------------------------------------------------------------------
\begin{abstract}
We investigate the magnetization process for a quasi-one-dimensional 
$S=1$ antiferromagnet with bond alternation.
By combining the density matrix renormalization group method
with the interchain mean-field theory,
we discuss how the interchain coupling affects the magnetization curve.
It is found that the width of the magnetization plateau is considerably 
reduced upon introducing the interchain coupling.
We obtain the phase diagram in a magnetic field.
The effect of single-ion anisotropy is also addressed. 
\end{abstract}


%%%\kword
%%%{$S=1$ spin chain, Haldane gap, DMRG, magnetization plateau }
%%%\begin{document}
%%%\sloppy
\maketitle

%%%%%%%%%%%%%%%%%%%%%%%%%%%%%%%%%%%%%%%%%%%%%%%%%%%%%%%%%%%%%%%%%%%
%%%%%%%%%%%%%%%%%%%%%%%%%%%%%%%%%%%%%%%%%%%%%%%%%%%%%%%%%%%%%%%%%%%
%%%%%%%%%                   　　　　  %%%%%%%%%%%%%%%%%%%%%%%%%%%%%
%%%%%%%%% 1.  Introduction  　　　　  %%%%%%%%%%%%%%%%%%%%%%%%%%%%%
%%%%%%%%%                   　　　　  %%%%%%%%%%%%%%%%%%%%%%%%%%%%%
%%%%%%%%%%%%%%%%%%%%%%%%%%%%%%%%%%%%%%%%%%%%%%%%%%%%%%%%%%%%%%%%%%%
%%%%%%%%%%%%%%%%%%%%%%%%%%%%%%%%%%%%%%%%%%%%%%%%%%%%%%%%%%%%%%%%%%%

\section{Introduction}

Low-dimensional quantum spin systems with the spin gap
have attracted much interest recently. 
In particular, the Haldane system with integer spin　
has been studied extensively. \cite{Haldane} 
The origin of the Haldane spin-gap phase   is now 
understood well in terms of the 
valence-bond-solid (VBS) picture. \cite{VBS} 
Theoretical investigations have been extended　to more realistic systems, including
{\it e.g.}  the bond alternation, 
\cite{Affleck,Singh,Yamamoto,Yasuda,Kim,Koga,Kato,Totsuka,Tone_1,Hida_1}
the interchain coupling, \cite{Yasuda,Kim,Koga,Sakai_0,Sakai_1}
the single-ion anisotropy,\cite{Tone_1,Hida_1,Sakai_0,Sakai_1,Sakai_2,Dterm} 
etc.  

In particular, the interchain coupling, which may
 trigger the quantum phase transition to a magnetically ordered phase, 
 is indispensable to explain the ground state of some 
$S=1$ compounds
such as $\rm CsNiCl_3$,\cite{CsNiCl3} $\rm NENP$,\cite{NENP} 
$\rm PbNi_2V_2O_8$,\cite{ANi2V2O8} $\rm SrNi_2V_2O_8$.\cite{ANi2V2O8}
Theoretically,  Sakai and Takahashi\cite{Sakai_0}  studied 
the effects of the interchain coupling on the Haldane system
by combining the exact diagonalization method
with the interchain mean-field theory,  and thus 
obtained the phase diagram for the quasi-one-dimensional
(Q1D) system. More quantitative treatments were later given 
by the series expansion method \cite{Koga}
and also  by the quantum Monte Carlo simulations. \cite{Yasuda,Kim}

Some striking phenomena for the quantum 
phase transition found under an applied magnetic field 
 have further stimulated the study on such Q1D spin systems. Most 
interesting may be
 the plateau formation in the magnetization curve at some fractional 
value of the full moment. Also, a long-range order induced 
by a magnetic field gives  another interesting 
paradigm of the quantum phase transition, which has been observed 
in some materials such as NDMAZ, \cite{NDMAZ} 
NDMAP, \cite{NDMAP} NTEAP \cite{NTEAP} and NTENP. \cite{NTENP,Narumi_2}

In this paper, we study magnetic properties of 
a Q1D $S=1$ Heisenberg antiferromagnet with bond alternation
by combining the density matrix renormalization group (DMRG) with 
the interchain mean-field theory.\cite{Sakai_0,Sakai_1,MF1/2}
We clarify how the interchain coupling affects the
magnetization curve with particular emphasis on the 
properties around the plateau. 
The effect of single-ion anisotropy is also discussed,
which  sometimes plays an essential role 
to understand low-energy properties of  Q1D 
$S=1$  compounds.

This paper is organized as follows. 
In Sec. II, we introduce the model Hamiltonian 
for a Q1D $S=1$ system, and 
 briefly outline the numerical procedure.
In Sec. III, we present the results, and 
discuss magnetic properties of the Q1D system.
Furthermore the 
effect of single-ion anisotropy is discussed in Sec. IV. 
It is shown that the magnetization curve exhibits distinct
behaviors depending on the 
relative direction between single-ion anisotropy
and an applied magnetic field. 
Brief summary is given in Sec. V.

%%%%%%%%%%%%%%%%%%%%%%%%%%%%%%%%%%%%%%%%%%%%%%%%%%%%%%%%%%%%%%%%%%%
%%%%%%%%%%%%%%%%%%%%%%%%%%%%%%%%%%%%%%%%%%%%%%%%%%%%%%%%%%%%%%%%%%%
%%%%%%%%%                      　　　　 %%%%%%%%%%%%%%%%%%%%%%%%%%%
%%%%%%%%% 2  　model and method　　　　 %%%%%%%%%%%%%%%%%%%%%%%%%%%
%%%%%%%%%                      　　　　 %%%%%%%%%%%%%%%%%%%%%%%%%%%
%%%%%%%%%%%%%%%%%%%%%%%%%%%%%%%%%%%%%%%%%%%%%%%%%%%%%%%%%%%%%%%%%%%
%%%%%%%%%%%%%%%%%%%%%%%%%%%%%%%%%%%%%%%%%%%%%%%%%%%%%%%%%%%%%%%%%%%

\section{ Model and Numerical Method}

In order to study a Q1D $S=1$ antiferromagnet in an applied magnetic
field $H$,  we consider the following Hamiltonian,
%%%%%%%%%%%%%%%%%%%%%%%%%%%%%%%%%%%%%%%%%%%%%%%%%%%%%%%%%%%%%%%%%%%%%%%%%%
%\begin{eqnarray}
%{\cal H}&=&{\cal H}_{intra}+{\cal H}_{inter}
%\cr
%{\cal H}&=&J
%      \sum_{i,\mbox{\boldmath$r$}}
%      \{1-(-1)^{i}\delta\}
%      \mbox{\boldmath$S$}_{i,\mbox{\boldmath$r$}}
%      \cdot\mbox{\boldmath$S$}_{i+1,\mbox{\boldmath$r$}}
%   \cr
%      +D\sum_{i,\mbox{\boldmath$r$}}
%      (S^{\alpha}_{i,\mbox{\boldmath$r$}})^2
%  \cr
%    &+&J'\sum_{ i,\mbox{\boldmath$r$},\mbox{\boldmath$\mu$} }
%%    \sum_{ \mbox{\boldmath$\alpha$}=1 }^{z}
%      \mbox{\boldmath$S$}_{i,\mbox{\boldmath$r$}}
%      \cdot\mbox{\boldmath$S$}
%      _{i,\mbox{\boldmath$r$}+\mbox{\boldmath$\mu$}}
%
%        -H\sum_{i,\mbox{\boldmath$r$}}
%        S^{z}_{i,\mbox{\boldmath$r$}}, 
%\label{3DHM}
%\end{eqnarray}
%%%%%%%%%%%%%%%%%%%%%%%%%%%%%%%%%%%%%%%%%%%%%%%%%%%%%%%%%%%%%%%%%%%%%%%%%%
%%%%%%%%%%%%%%%%%%%%%%%%%%%%%%%%%%%%%%%%%%%%%%%%%%%%%%%%%%%%%%%%%%%%%%%%%%
\begin{eqnarray}
{\cal H}&=&J
      \sum_{j,i}
      \{1-(-1)^{i}\delta\}
      \mbox{\boldmath$S$}_{j,i}
      \cdot\mbox{\boldmath$S$}_{j,i+1}
%%   \cr
%      +D\sum_{j,i}
%      (S^{z}_{j,i})^2
  \cr
   && +J'\sum_{ <j,j'>, i }
%%    \sum_{ \mbox{\boldmath$\alpha$}=1 }^{z}
      \mbox{\boldmath$S$}_{j,i}
      \cdot\mbox{\boldmath$S$}_{j',i}
        -H\sum_{j,i}  S^{z}_{j,i}, 
\label{3DHM}
\end{eqnarray}
%%%%%%%%%%%%%%%%%%%%%%%%%%%%%%%%%%%%%%%%%%%%%%%%%%%%%%%%%%%%%%%%%%%
where $\mbox{\boldmath$S$}_{j,i}$ is the $S=1$ spin operator at the $i$-th site 
in the $j$-th chain.  The summation with 
$<j,j'>$ is taken over all the pairs of the nearest-neighbor chains. 
Here $\delta$ is the bond alternation parameter and
$J(J')$ the intra-chain (inter-chain) antiferromagnetic coupling.
We set $J=1$ and $g\mu_B=1$, for convenience.

We discuss magnetic properties of the above  Q1D 
spin model by combining DMRG  \cite{White,Nishino} with 
the interchain mean-field theory.\cite{Sakai_0,Sakai_1,MF1/2}
To deal with the effects of the interchain coupling 
in the presence of a magnetic field, 
we introduce two kinds of mean fields as $<S_i^x>\sim (-1)^{i} m_s$ 
and  $<S_i^z>\sim m_u$, where $m_s (m_u)$ is 
the staggered (uniform) moment induced by the interchain coupling and 
the applied magnetic field. 
We thus end up with the 1D Hamiltonian, 
%%%%%%%%%%%%%%%%%%%%%%%%%%%%%%%%%%%%%%%%%%%%%%%%%%%%%%%%%%%%%%%%%%%%%%%%%%
\begin{eqnarray}
{\cal H}_{MF}&=&
      \sum_{i}\{ 1-(-1)^{i}\delta \}
      \mbox{\boldmath$S$}_{i}\cdot\mbox{\boldmath$S$}_{i+1}
   \cr
%      &+& 
%D\sum_{i} (S^{\alpha}_{i})^2
      &&   -\left(H-h_u\right)\sum_{i} S^{z}_{i} 
        -h_{s}\sum_{i} (-1)^{i} S_{i}^{x}.
\label{MFHM}
\end{eqnarray}
%%%%%%%%%%%%%%%%%%%%%%%%%%%%%%%%%%%%%%%%%%%%%%%%%%%%%%%
Here $h_s$ and $h_u$ are the effective internal fields, which 
are defined as $h_s=z_{\perp}J'm_s$ and $h_u=z_{\perp}J'm_u$ 
respectively, where $z_{\perp}$ is the number of the adjacent chains.
The staggered (uniform) magnetization per site $m_s$ ($m_u$) is 
written down as
%%%%%%%%%%%%%%%%%%%%%%%%%%%%%%%%%%
\begin{eqnarray}
m_s&=&\phi_s\left( h_s, h_u, \delta, H \right),\\
m_u&=&\phi_u\left( h_s, h_u, \delta, H \right),
\end{eqnarray}
%%%%%%%%%%%%%%%%%%%%%%%%%%%%%%%%%%%%%%%%
where the functions $\phi_s$ and $\phi_u$ can be determined
from the magnetization for the effective spin chain with both of the 
uniform and staggered fields.
Since the mean-field Hamiltonian 
(\ref{MFHM}) is given as a function of $H-h_u$,
it is convenient to rewrite the staggered magnetization as
%%%%%%%%%%%%%%%%%%%%%%%%%%%%%%%%%%%%%%%%%%%%%%%%%%%%%%%%
\begin{eqnarray}
m_s&=&\phi_s (h_s, \delta, \tilde{H} ),
\label{SF-EQ}
\end{eqnarray}
%%%%%%%%%%%%%%%%%%%%%%%%%%%%%%%%%%%%%%%%%%%%%%%%
where the effective field is defined as $\tilde{H}=H-h_u$.
By solving the equation  with the self-consistent condition
 $h_s=z_{\perp}J'm_s$,
we can discuss the quantum phase transition from the 
 spin-gap phase to a  magnetically ordered phase. Namely,
if this self-consistent equation has a finite $m_s$,
a long-range order is induced by the interchain coupling.
Otherwise, the staggered field induced by the interchain coupling 
 becomes irrelevant, and thereby
the spin-gap phase still persists against a magnetically
ordered phase.  The critical value $J'_c$ for the phase transition 
 is given in terms of  the zero-field staggered susceptibility as,
%%%%%%%%%%%%%%%%%%%%%%%%
\begin{eqnarray}
\frac{1}{z_{\perp}J'_c}
\equiv\left. \frac{\partial\phi_s}{\partial h_s}\right|_{h_s\rightarrow 0}.
\end{eqnarray}
%%%%%%%%%%%%%%%%%%%%%%%%
%%Thus by evaluating the function $\phi_s$ around 
%%the zero field carefully.
This completes the formulation of the interchain mean-field 
treatment.

We conclude this section by giving an explicit form of 
the staggered magnetization $\phi_s$ for the single chain case,
which will play a key role in the following analysis.
We calculate  $\phi_s$ by means of the infinite DMRG method, which
enables us to treat 1D spin systems even when
 the total $S^z$ is not a conserved quantity. 
%%%%%%%%%%%%%%%%%%%%%%%%%%%%%%%%%%%%%%%%%%%%%%
\begin{figure}[htb]
\begin{center}
\vspace{-0.0cm}
\hspace{-0.5cm}
\leavevmode \epsfxsize=80mm 
\epsffile{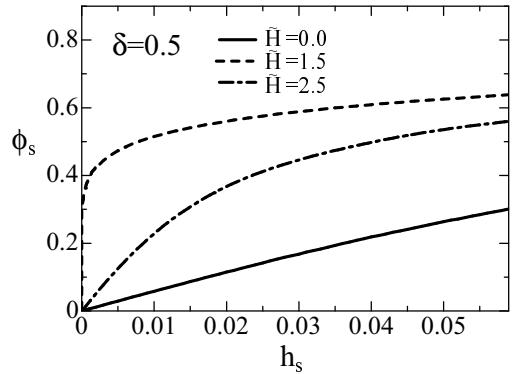}
\vspace{-6.0cm}
\end{center}
\caption{Staggered magnetization $\phi_s$ as a function of the staggered
field $h_s$ for an isolated spin chain with bond alternation $\delta=0.5$.
The solid, broken and dot-dashed lines are the magnetization curves
for $\tilde{H}=0.0, 1.5$ and $2.5$, respectively.
%Thin lines are $m_s=h_s/z_{\perp}J'$.  
}
\label{hx_Sx-FIG}
\end{figure}
%%%%%%%%%%%%%%%%%%%%%%%%%%%%%%%%%%%%%%%%%%%%%%%%%%%%%%%%%%%%%%%%%%%
In Fig. \ref{hx_Sx-FIG}, we summarize 
characteristic profiles of the staggered magnetization 
$\phi_s$ with $\tilde{H}$ being fixed.
When $\tilde{H}=0$, the system has the triplet-excitation gap 
over the dimer-singlet ground state stabilized
 by the bond alternation $(\delta=0.5)$.
In fact, the magnetization curve 
in staggered fields has a finite derivative at the origin, as seen
from the solid line in Fig. \ref{hx_Sx-FIG}, implying
that the dimerized ground state is stable up to a certain critical value.
%%%%%%%$(z_{\perp}J'_c\simeq0.17)$.
The introduction of an external field  decreases the spin gap,
and eventually triggers the quantum  phase transition 
to a gapless phase (so-called Tomonaga-Luttinger 
phase) at  $\tilde{H}=1.5$, which is accompanied by
the divergence of the spin
susceptibility shown as the dashed line in Fig. \ref{hx_Sx-FIG}.
 It is seen that the
staggered susceptibility takes a finite value again
in the large field region ($\tilde{H}=2.5$), implying that
 the spin gap phase with the magnetization plateau
is stabilized, for which a finite interchain coupling 
%%%%$z_{\perp}J'_c\simeq 0.037$ 
is necessary to destroy the spin-gap state.

%%%%%%%%%%%%%%%%%%%%%%%%%%%%%%%%%%%%%%%%%%%%%%%%%%%%%%%%%%%%%%%%%%%%%
%%%%%%%%%%%%%%%%%%%%%%%%%%%%%%%%%%%%%%%%%%%%%%%%%%%%%%%%%%%%%%%%%%%%%
%%%%%%%%%%%%%%%%%%%%                         　　　%%%%%%%%%%%%%%%%%%
%%%%%%%%%%%%%%%%%%%%   3. RESULTS            　　　%%%%%%%%%%%%%%%%%%
%%%%%%%%%%%%%%%%%%%%                         　　　%%%%%%%%%%%%%%%%%%
%%%%%%%%%%%%%%%%%%%%%%%%%%%%%%%%%%%%%%%%%%%%%%%%%%%%%%%%%%%%%%%%%%%%%
%%%%%%%%%%%%%%%%%%%%%%%%%%%%%%%%%%%%%%%%%%%%%%%%%%%%%%%%%%%%%%%%%%%%%
\section{Effects of the interchain coupling}

%%%%%%%%%%%%%%%%%%%%%%%%%%%%%%%%%%%%%%%%%%%%%%%%%%%%%%%%%%%%%%%%%%%
%%%%%%%%% 3.1　 D=0          　　　　 %%%%%%%%%%%%%%%%%%%%%%%%%%%%%
%%%%%%%%%%%%%%%%%%%%%%%%%%%%%%%%%%%%%%%%%%%%%%%%%%%%%%%%%%%%%%%%%%%

We now discuss how the interchain coupling affects the 
magnetization process in the Q1D system, following the 
procedure outlined above.  In the following
we will explicitly use  the bare external field $H$  
instead of $\tilde H$ ($H=\tilde{H}+h_u$).

\begin{figure}[thb]
\begin{center}
\vspace{-0.0cm}
\hspace{-0.5cm}
\leavevmode \epsfxsize=80mm 
\epsffile{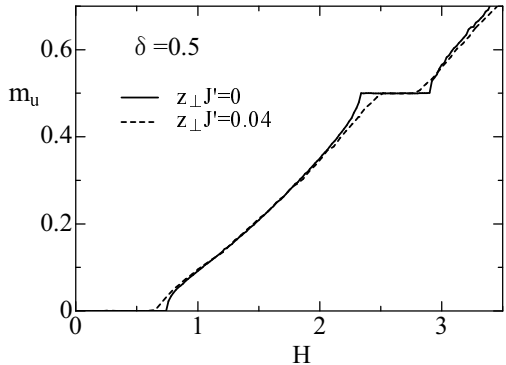}
\vspace{-6.0cm}
\end{center}
\caption{ Magnetization curves for $\delta=0.5$ as a function 
of the applied field $H$.
The solid and broken lines are the results calculated for 
an isolated chain ($z_{\perp}J'=0$) and 
 weakly coupled chains ($z_{\perp}J'=0.04$), respectively.  
}
\label{mag_D0-FIG} 
\end{figure}

In Fig. \ref{mag_D0-FIG}, we show the magnetization curve calculated for
the Q1D system, which is compared with that for the pure 1D case ($J'=0$).
%%%%%%%%%%%%%%%%%%%%%%%%%%%%%%%%%%%%%%%%%%%%%%%%%%%%%%%%%%%%%%%%%
%%%%%%%%%%%%%%%%%%%%%%%%%%%%%%%%%%%%%%%%%%%%%%%%%%%%%%%%%%%%%%%
Note that the ground-state for the 1D case is 
either in the singlet-dimer phase ($\delta >0.25$) or 
in the Haldane phase ($\delta <0.25$).
We show the magnetization for the dimer-singlet phase
 ($\delta=0.5$) as an example, since that for the Haldane 
phase exhibits a similar behavior.
Let us begin with  the magnetization 
for the 1D case. Since the system is in the singlet-dimer 
phase with the spin excitation gap, the uniform magnetization 
does not appear up to the critical field $H_{c0}$.
Beyond the critical field, the magnetization begins to increase and
the system enters the Tomonaga-Luttinger liquid phase,
which is immediately driven to a magnetically ordered phase 
upon introducing the interchain coupling.
Further an increase in the field drives the system to another 
plateau state at $m_u=1/2$.  This spin-gap phase
with  $m_u=1/2$ is expected to be stable against
the introduction of the interchain coupling.
However, it should be noticed that  the spin-gap vanishes 
at  both ends of the magnetization plateau,
so that the width of the plateau should be
considerably reduced by the interchain coupling.
We can see that
this is indeed the case for the results in Fig.\ref{mag_D0-FIG}.

\begin{figure}[htb]
\begin{center}
\vspace{-0.cm}
\hspace{-0.5cm}
\leavevmode \epsfxsize=80mm 
\epsffile{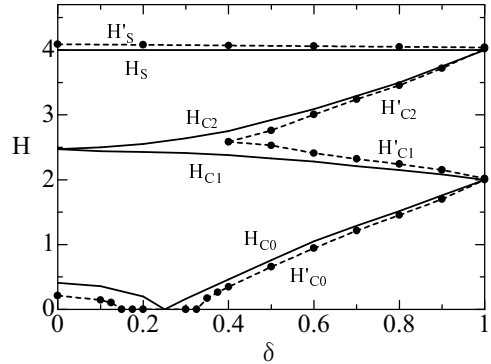}
\vspace{-5.5cm}
\end{center}
\caption{Phase diagram for the Q1D spin system.
The solid and broken lines represent the phase diagram for
an isolated chain ($z_{\perp}J'=0$) and 
weakly coupled chains ($z_{\perp}J'=0.04$), respectively.
The uniform magnetization takes the quantized value 
at $m_u=0$ in the region 
from $0$ to $H_{c1}$ ($H'_{c1}$), 
$m_u=1/2$ in the region from $H_{c2}$ ($H'_{c2}$) to $H_{c3}$ ($H'_{c3}$), 
and $m_u=1$ above $H_s$ ($H'_s$). In other regions, 
the magnetization varies  continuously.
}
\label{PD_D0-FIG}
\end{figure}

To see the above behavior in more detail, we show the phase diagram
for the Q1D spin system with 
the interchain coupling
$z_{\perp}J'=0$ and $0.04$ in Fig. \ref{PD_D0-FIG}.
%%%%%%%%%%%%%%%%%%%%%%%%%%%%%%%%%%%%%%%%%%%%%%%%%%%%%%%%%%%%%%%%%%%%%%%%%%%%
%%%%%%%%%%%%%%%%%%%%%%%%%%%%%%%%%%%%%%%%%%%%%%%%%%%%%%%%%%%%%%%%%%%
Note that our phase diagram for the isolated chain ($z_{\perp}J'=0$) 
agrees fairly well with that already obtained by the exact 
diagonalization method. \cite{Tone_1}
Let us consider how the interchain coupling affects the  plateau
region with $m_u=1/2$, which is stabilized by the bond alternation.
As seen from Fig. \ref{PD_D0-FIG}, 
the region of this phase is extended
with increasing $\delta$.  On the other hand, when
the interchain coupling is introduced, the quantum 
fluctuation stabilizing the 
plateau is suppressed, and thus the region for
the $m_u=1/2$ plateau is reduced.

We note here that the phase diagram obtained here 
is qualitatively valid except for an extreme case, $\delta\sim 1$, 
where the system is approximately described by 
an assembly of isolated dimers for $J'=0$.  
When a finite interchain coupling 
$J'$ is introduced for such almost isolated  
dimers, the system first forms the 
ladders rather than the chains for two-dimension.  
In this case, our approach based on the interchain mean-field theory may not
be sensible.  To treat  the critical phenomena
in this region, we need more precise treatment
 beyond mean-field theory, 
which is now under consideration.

Before closing this section, a brief comment is in order for 
 the magnetization process in the Q1D Haldane system 
with a ferromagnetic interchain coupling, $J'<0$.
By rewriting the effective Hamiltonian,
the critical points of the system with the ferromagnetic 
interchain couplings 
are given as, 
%%%%%%%%%%%%%%%%%%%%%%%%
\begin{eqnarray}
&&  H'_{c0}(J'=-|J'|)=H'_{c0}(J'=|J'|) 
\label{F1}  \\
&&  H'_{c1,c2}(J'=-|J'|)=H'_{c1,c2}(J'=|J'|)-z_{\perp}|J'|  
\label{F2}  \\
&&  H'_{s}(J'=-|J'|)=H'_{s}(J'=|J'|)-2z_{\perp}|J'|.  
\label{F3}
\end{eqnarray}
%%%%%%%%%%%%%%%%%%%%%%%%
As easily seen from  these formulae, 
the width of the plateau $H'_{c2}-H'_{c1}$ does not depend on
the sign of $J'$ in the framework of our mean-field treatment.

%%%%%%%%%%%%%%%%%%%%%%%%%%%%%%%%%%%%%%%%%%%%%%%%%%%%%%%%%%%%%%%%%%%%%%%%%%%%
%\begin{figure}[htb]
%\begin{center}
%\vspace{-0.cm}
%\hspace{-0.5cm}
%\leavevmode \epsfxsize=80mm 
%\epsffile{H_Jp_D0.eps}
%\vspace{-4.0cm}
%\end{center}
%\caption{ $H-z_{\perp}J'$ phase diagram about 
%(a) $M=0$ and (b) $M=1/2$ for $\delta=0.5$ and $D=0$.  
%We estimate $z_{\perp}J'_c$  at $h_s=0.001, 0.002, 0.003$. 
%Solid curves are our fitted lines.
%}
%\label{H_Jp-FIG}
%\end{figure}
%%%%%%%%%%%%%%%%%%%%%%%%%%%%%%%%%%%%%%%%%%%%%%%%%%%%%%%%%%%%%%%%%%%

%%%%%%%%%%%%%%%%%%%%%%%%%%%%%%%%%%%%%%%%%%%%%%%%%%%%%%%%%%%%%%%%%%%
%%%%%%%%% 3.2　 D=0.5        　　　　 %%%%%%%%%%%%%%%%%%%%%%%%%%%%%
%%%%%%%%%%%%%%%%%%%%%%%%%%%%%%%%%%%%%%%%%%%%%%%%%%%%%%%%%%%%%%%%%%%

\section{Effects of single-ion anisotropy}

In this section, we discuss the effect of single-ion anisotropy
 with the Hamiltonian, ${\cal H}_D=D\sum_i(S^{\alpha}_i)^2$.
We consider two cases of anisotropy with respect
to the direction of an applied magnetic field, i.e. $D\perp H$ ($\alpha=y$)
 and $D\parallel H$ ($\alpha=z$).
By choosing the $x$-axis as the  axis for the 
magnetic order, we 
 discuss the magnetic properties of the Q1D system.

%%%%%%%%%%%%%%%%%%%%%%%%%%%%%%%%%%%%%%%%%%%%%%%%%%%%%%%%%%%%%%%%%%%%%%%%%%%%
\begin{figure}[htb]
\begin{center}
\vspace{-0.cm}
\hspace{-0.5cm}
\leavevmode \epsfxsize=80mm 
\epsffile{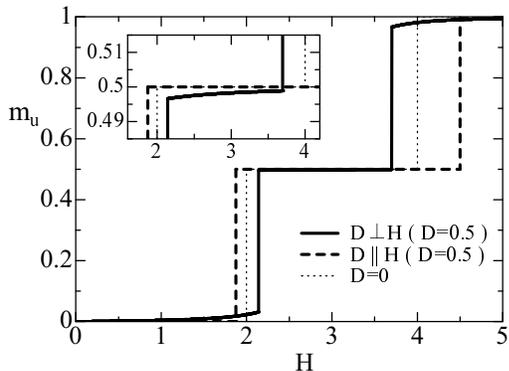}
\vspace{-5.5cm}
\end{center}
\caption{Magnetization curves for the isolated dimer system ($\delta=1$). 
For $D \parallel H (D=0.5)$, $D\perp H (D=0.5)$ and $D=0$.
The inset shows the behavior around the plateau $m_u=1/2$. 
}
\label{2spin-FIG}
\end{figure}
%%%%%%%%%%%%%%%%%%%%%%%%%%%%%%%%%%%%%%%%%%%%%%%%%%%%%%%%%%%%%%%%%%%

To begin with, let us consider a simple system in the 
isolated-dimer limit  with $(\delta, J')=(1, 0)$, which provides us
with some insight into the role of single-ion anisotropy.
The magnetization curves for $D=0.5$ are 
shown as the broken line ($D\parallel H$) and 
the solid line ($D\perp H$) in Fig.\ref{2spin-FIG}.
Note that the magnetization shows jumps in
a magnetic field, reflecting the isolated dimer-singlet
ground state
When $D\parallel H$, the plateaus with $m_u=0$, $1/2$ and $1$ 
appear in the magnetization process clearly.
On the other hand, in the case of $D\perp H$, the plateau
is not exactly flat but is smeared  by the Van Vleck contribution,
as seen from the enlarged picture in the  inset.
This is because  the total $S^z$ is not a conserved quantity  
for the case of $D\perp H$ even for $h_s=0$, 
which is in contrast to the case of $D\parallel H$. 
We also find that the width of the $m_u=1/2$ plateau is 
increased (decreased) upon introducing anisotropy,
  $D\parallel H$ ($D\perp H$).  Therefore,  
 the observation of the plateau structure may be 
somewhat difficult for $D\perp H$ in generic Q1D spin
systems, because of shrinking 
of the width as well as smearing due to the Van Vleck term.

%%%%%%%%%%%%%%%%%%%%%%%%%%%%%%%%%%%%%%%%%%%%%%%%%%%%%%%%%%%%%%%%%%%%%%%%%%%%
\begin{figure}[htb]
\begin{center}
\vspace{-0.cm}
\hspace{-0.5cm}
\leavevmode \epsfxsize=80mm 
\epsffile{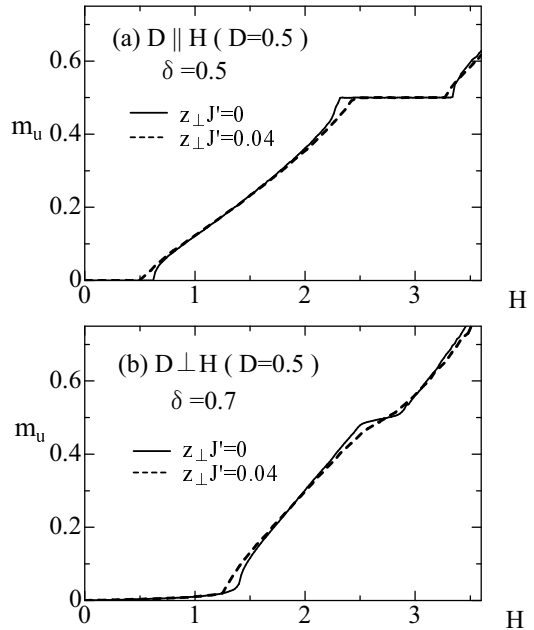}
\vspace{-2.5cm}
\end{center}
\caption{ Magnetization curves for the Q1D system:
(a) $D \parallel H (D=0.5)$ and $\delta=0.5$, and
(b) $D\perp H (D=0.5)$ and $\delta=0.7$.
The solid and broken lines show the results for $z_{\perp}J'=0$ and 
$0.04$, respectively.
}
\label{MAG_D05-FIG}
\end{figure}
%%%%%%%%%%%%%%%%%%%%%%%%%%%%%%%%%%%%%%%%%%%%%%%%%%%%%%%%%%%%%%%%%%%

Keeping the above properties in mind, we now consider 
the magnetization process for the pure 1D  as well as the Q1D
systems.  The numerical results are summarized in 
Fig. \ref{MAG_D05-FIG}.  
Shown in Fig. \ref{MAG_D05-FIG}(a) are the 
magnetization curves for $D \parallel H (D=0.5)$ and $\delta=0.5$,
Since the presence of anisotropy, $D \parallel H$,  has a tendency to
stabilize the  $m_u=1/2$ plateau, as mentioned above, thus
increasing the plateau width  compared with the
isotropic case $D=0$ (see Fig. \ref{mag_D0-FIG}).
It is also seen that the plateau state is rather stable
against the introduction of the interchain coupling.

\begin{figure}[htb]
\begin{center}
\vspace{-0.cm}
\hspace{-0.5cm}
\leavevmode \epsfxsize=80mm 
\epsffile{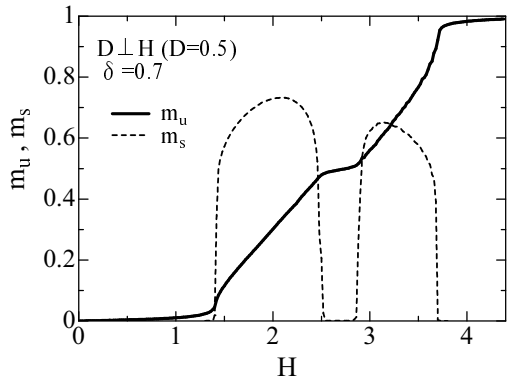}
\vspace{-6.0cm}
\end{center}
\caption{ Uniform and staggered magnetizations 
for an isolated spin chain when $D\perp H$ ($D=0.5$).
}
\label{H_ms-FIG}
\end{figure}

In Fig. \ref{MAG_D05-FIG} (b), we show the magnetization curves 
for the case of $D\perp H$ with bond alternation $\delta=0.7$. 
%%We note that the $m_u\sim 1/2$ plateau disappears
%%beyond the critical interchain interaction $z_{\perp}J'=0.04$. 
 In contrast to the above $D \parallel H$ case, the plateau 
is indeed obscured
both for the 1D (solid line) and Q1D systems (dashed line),
which is due to the Van Vleck contribution.
Therefore, in order to obtain the phase transition point
correctly, it is necessary to  consider  another proper 
quantity characterizing the transition clearly. For this purpose,
we evaluate the staggered magnetization in the $x$-axis.
The calculated results for a single-chain
system are plotted  
(broken line) together with the uniform magnetization (solid line)
 in Fig. \ref{H_ms-FIG}.
We note that a negligibly small staggered field 
$h_s=1.0\times 10^{-5}$ is added 
to stabilize the spontaneous staggered magnetization.
It is seen that although the plateaus at $m_u=0, 1/2$ and $1$ 
are vague, the phase transition point can be determined 
clearly at which the staggered magnetization vanishes. 
This is used to determine the phase diagram for the
case of $D\perp H$.

%%%%%%%%%%%%%%%%%%%%%%%%%%%%%%%%%%%%%%%%%%%%%%%%%%%%%%%%%%%%%%%%%%%%%%%%%%%%
%%%%%%%%%%%%%%%%%%%%%%%%%%%%%%%%%%%%%%%%%%%%%%%%%%%%%%%%%%%%%%%%%%%

\begin{figure}[htb]
\begin{center}
\vspace{-0.cm}
\hspace{-0.5cm}
\leavevmode \epsfxsize=80mm 
\epsffile{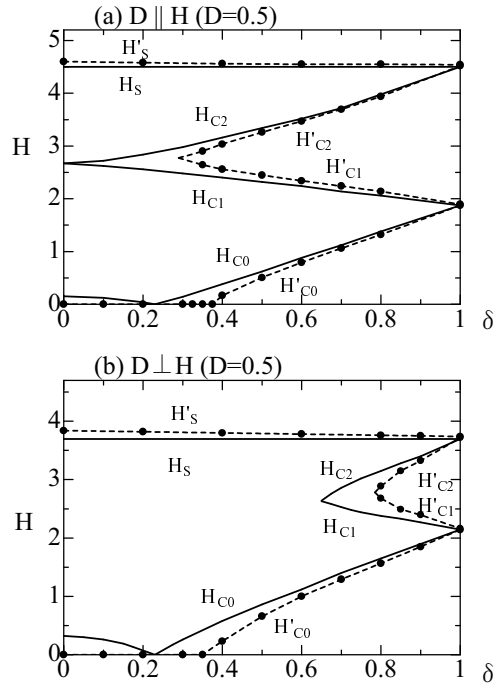}
\vspace{-1.5cm}
\end{center}
\caption{Phase diagrams for the Q1D quantum spin chain with 
single-ion anisotropy.
The solid and broken lines are the phase boundaries for an
isolated chain ($z_{\perp}J'=0$) and 
 weakly coupled chains ($z_{\perp}J'=0.04$), respectively.
The meanings of the critical fields and the 
corresponding phases are as in Fig. \ref{PD_D0-FIG}.
 }
\label{PD_D05-FIG}
\end{figure}

Finally, we show the $\delta$-$H$ phase diagram for 
$D\parallel H$ and $D\perp H$
in Fig. \ref{PD_D05-FIG} (a) and (b).
In these figures, the solid and broken lines 
give the phase diagrams  for 
an isolated 1D chain and weakly coupled chains, respectively.
We note that  when 
 $J'=H=0$, the quantum phase transition between
the Haldane phase and the singlet-dimer phase
occurs at the critical point 
$\delta _c\simeq 0.23$ for $ D=0.5$. \cite{Tone_1,Hida_1}
In the case of $D \parallel H$, the effects of the interchain couplings are 
analogous 
to those for the $D=0$ case (see Fig. \ref{PD_D0-FIG}).
On the other hand, when the system has anisotropy in the $y$-direction, 
i.e. $D\perp H$, the phase diagram exhibits a quite different 
feature in contrast to the $D\parallel H$ case.  In particular,
the $m_u=1/2$ plateau region is reduced considerably both for
the pure 1D as well as the Q1D systems.

%%%%%%%%%%%%%%%%%%%%%%%%%%%%%%%%%%%%%%%%%%%%%%%%%%%%%%%%%%%%%%%%%%%%%%%%%%%%
%%%%%%%%%%%%%%%%%%%%%%%%%%%%%%%%%%%%%%%%%%%%%%%%%%%%%%%%%%%%%%%%%%%

%%%%%%%%%%%%%%%%%%%%%%%%%%%%%%%%%%%%%%%%%%%%%%%%%%%%%%%%%%%%%%%%%%%
%%%%%%%%% 4. 　 SUMMARY 　　　　　　  %%%%%%%%%%%%%%%%%%%%%%%%%%%%%
%%%%%%%%%　　　　　　　　　　　　　　 %%%%%%%%%%%%%%%%%%%%%%%%%%%%%
%%%%%%%%%%%%%%%%%%%%%%%%%%%%%%%%%%%%%%%%%%%%%%%%%%%%%%%%%%%%%%%%%%%
\section{Summary}

We have studied the magnetization process
for a Q1D $S=1$ antiferromagnet with bond alternation.
By combining the density matrix renormalization group method 
with the interchain mean-field treatment,
the magnetic properties have been discussed. 
It has been  shown that the introduction of the interchain coupling enhances 
the antiferromagnetic correlation, thereby reducing
the width of the plateau in the magnetization curve.

We have also discussed the effects of single-ion anisotropy, 
which sometimes plays an important role for understanding experimental
findings in $S=1$ spin systems.
It has been clarified that the magnetization 
curves exhibit distinct behaviors depending on the relative 
direction between  single-ion anisotropy 
and an applied field.  In particular, we have 
found that the magnetization plateau in the case 
of $D\perp H$ is rather unstable against the interchain coupling, 
and also suffers from the smearing effect
due to the Van Vleck contribution, making the experimental 
observation somewhat difficult.

%We think that the present study is useful to understand 
%the magnetization process in Q1D S=1 Heisenberg 
%antiferromagnet compounds.

%Recently the magnetic properties for the compound 
% [Ni(333-tet)($\mu$ -NO$_2$)](ClO$_4$) are reported 
% by Y. Narumi {\it et. al}.\cite{Narumi2} 
%It is shown that the interchain coupling plays an important role 
%to understand the magnetization process. 

%comment ZM-interaction is possible ?

%%%%%%%%%%%%%%%%%%%%%%%%%%%%%%%%%%
\section*{Acknowledgements}
%%%%%%%%%%%%%%%%%%%%%%%%%%%%%%%%%%%%%
We would like to thank Y. Narumi and K. Kindo for valuable discussions. 
This work was partly supported by a Grant-in-Aid from the Ministry 
of Education, Science, Sports and Culture of Japan. 
A part of computations was done at the Supercomputer Center at the 
Institute for Solid State Physics, University of Tokyo
and Yukawa Institute Computer Facility. 
A. Kawaguchi was supported by 
Japan Society for the Promotion of Science. 

%%%%%%%%%%%%%%%%%%%%%%%%%%%%%%%%%%%%%%%%%%%%%%%

%\clearpage
%%%%%%%%%%%%%%%%%%%%%%%%%%%%%%%%%%%%%%%%%%%%%%%%%%%%%%%%%%%%%%%%%%%
%%%%%%%%%    　Reference  　　　　　  %%%%%%%%%%%%%%%%%%%%%%%%%%%%%
%%%%%%%%%　　　　　　　　　　　　　　 %%%%%%%%%%%%%%%%%%%%%%%%%%%%%
%%%%%%%%%%%%%%%%%%%%%%%%%%%%%%%%%%%%%%%%%%%%%%%%%%%%%%%%%%%%%%%%%%%

\end{document}